\documentclass[aps, prd, amsmath, floats, floatfix, twocolumn, nofootinbib, superscriptaddress,showpacs]{revtex4-1}
\usepackage{graphicx}
\usepackage{color}
\usepackage{latexsym}

\newcommand{\be}{\begin{eqnarray}}
\newcommand{\ee}{\end{eqnarray}}

\begin{document}

\title{Constraining the Kerr parameters via X-ray reflection spectroscopy}

\author{M. Ghasemi-Nodehi}
\affiliation{Center for Field Theory and Particle Physics and Department of Physics, Fudan University, 200433 Shanghai, China}

\author{Cosimo Bambi}
\email[Corresponding author: ]{bambi@fudan.edu.cn}
\affiliation{Center for Field Theory and Particle Physics and Department of Physics, Fudan University, 200433 Shanghai, China}
\affiliation{Theoretical Astrophysics, Eberhard-Karls Universit\"at T\"ubingen, 72076 T\"ubingen, Germany}

\date{\today}

\begin{abstract}
In a recent paper [Ghasemi-Nodehi \& Bambi, EPJC 76 (2016) 290], we have proposed a new parametrization for testing the Kerr nature of astrophysical black hole candidates. In the present work, we study the possibility of constraining the ``Kerr parameters'' of our proposal using X-ray reflection spectroscopy, the so-called iron line method. We simulate observations with the LAD instrument on board of the future eXTP mission assuming an exposure time of 200~ks. We fit the simulated data to see if the Kerr parameters can be constrained. If we have the correct astrophysical model, 200~ks observations with LAD/eXTP can constrain all the Kerr parameters with the exception of $b_{11}$, whose impact on the iron line profile is extremely weak and its measurement looks very challenging.
\end{abstract}

\pacs{04.70.-s, 97.60.Lf, 97.80.Jp}

\maketitle


\section{Introduction}

In 4-dimensional general relativity, the only stationary, axisymmetric, and asymptotically flat solution of the vacuum Einstein equations without spacetime singularities or closed time-like curves on or outside the event horizon is the Kerr metric~\cite{h1,h2}. It is remarkable that the spacetime metric around astrophysical black holes formed from gravitational collapse should be well approximated by the Kerr solution. After the formation of the black hole, initial deviations from the Kerr metric are quickly radiated away by the emission of gravitational waves~\cite{k1}. The equilibrium electric charge is soon reached, because of the highly ionized host environment around these objects, and its value cannot appreciably change the background metric~\cite{v2-k2a,v2-k2b,k2}. The presence of an accretion disk is normally negligible, because the mass of the disk is several orders of magnitude smaller than the mass of its black hole, and its impact on the background metric can be ignored~\cite{nt-model,k3}.

The metric around astrophysical black holes may deviate from the Kerr metric in the presence of new physics. Such a possibility is predicted in a number of different scenarios, ranging from alternative theories of gravity~\cite{np1} to the existence of exotic matter fields~\cite{np2} or macroscopic quantum gravity effects~\cite{np3,np4}. Recently, there has been a lot of work to study how to test the nature of astrophysical black holes and confirm the Kerr black hole hypothesis, with either electromagnetic radiation or gravitational waves; see, for instance, the recent reviews~\cite{r1,r2,r3,r4,r5}. The ``strategy'' to perform model-independent tests is to consider a spacetime more general than the Kerr solution and that includes the Kerr solution as a special case. Theoretical predictions are calculated in this general spacetime and then compared to observations to see if it is possible to constrain the spacetime metric around astrophysical black holes. There are several parametrized metrics in the literature, each of these with its advantages and disadvantages~\cite{t1,t2,t3,t4,t5,t6,t7}.

In Ref.~\cite{p1}, we proposed a new parametrization based on a number of ``Kerr parameters'', which are all 1 in the case of the Kerr metric and may be different from 1 in the presence of deviations from the predictions of general relativity. In the same paper, we studied the impact of these parameters on the shape of the apparent photon capture sphere, a measurement that might be possible in the near future, for instance with sub-mm very-long baseline interferometry facilities. Without a quantitative analysis, we found that the Kerr parameters $b_4$, $b_5$, $b_7$, and $b_{11}$ have a very weak impact on the shape of the apparent photon capture sphere, and presumably it will be very difficult to constrain their values. The other Kerr parameters can at least produce some clear effect on the shape of the apparent photon capture sphere, even if this does not necessarily  mean that it can be done with near future facilities.

In the present paper, we study the impact of the Kerr parameters on the shape of the iron line in the reflection spectrum. We want to understand whether it is possible to constrain the Kerr parameters via X-ray reflection spectroscopy. We simulate observations of a bright black hole binary with the LAD instrument on board of eXTP, a China-Europe X-ray mission which is currently scheduled to be launched in 2022. We assume an exposure time of 200~ks in all our simulations. Our results are quite promising. We find that all the Kerr parameters -- with the exception of $b_{11}$ -- can be potentially constrained using X-ray reflection spectroscopy. $b_{11}$ can unlikely be measured with this technique. $b_5$ and $b_8$ are more difficult to measure, so the resulting constraints are weaker or we need observations with a longer exposure time.

The content of the paper is as follows. In Section~\ref{s-p}, we briefly review the parametrization proposed in Ref.~\cite{p1}. In Section~\ref{s-r}, we study the iron K$\alpha$ lines expected in the reflection spectrum of accretion disks in our non-Kerr spacetimes. In Section~\ref{s-s}, we simulate 200~ks observations with LAD/eXTP. We fit the simulated observations with Kerr models to see whether it is possible to test the Kerr parameters of our parametrization. A summary and conclusions are in Section~\ref{s-c}. In the next sections, we will employ natural units in which $G_{\rm N} = c = 1$ and adopt a metric with the signature $(-+++)$.

\section{Parameterization of the Kerr metric \label{s-p}}

In the past 60~years, Solar System experiments have tested the Schwarzschild solution in the weak field limit. In the Parametrized Post-Newtonian (PPN) formalism, we write the most general static, spherically symmetric, and asymptotically flat metric as an expansion of $M/r$. In the so-called isotropic coordinates, the line element reads 
\be\label{eq-ppn}
ds^2 &=& - \left( 1 - \frac{2 M}{r} + \beta \frac{2 M^2}{r^2} + ... \right) dt^2 
\nonumber\\ &&
+ \left( 1+ \gamma \frac{2M}{r} + ... \right) \left( dx^2 + dy^2 + dz^2 \right) \, .
\ee
The term $2M/r$ in $g_{tt}$ is to recover the correct Newtonian limit, while the coefficients $\beta$ and $\gamma$ are introduced to parameterize our ignorance. From the comparison of theoretical predictions and observational data, we measure $\beta$ and $\gamma$. In general relativity, the only static and spherically symmetric solution of the vacuum Einstein equation is the Schwarzschild metric and, when it is written in isotropic coordinates, we see that $\beta = \gamma = 1$. Current measurements constrain $\beta$ and $\gamma$ to be 1 with a precision, respectively, of the order of $10^{-4}$ and $10^{-5}$~\cite{will}. This confirms the Schwarzschild solution in the weak field limit within the precision of current observations.

A similar strategy may be employed to test the Kerr metric in the strong gravity region of astrophysical black holes. However, this is not so easy. Now it is not possible to perform an expansion in $M/r$, because this is not a small parameter any longer. This leads to having an infinite number of possible deviations from the Kerr metric, and there is not a model-independent way to have a hierarchical structure of the deformations of the metric. Different authors have proposed different parametrized metrics, see~\cite{t1,t2,t3,t4,t5,t6,t7}.

In Ref.~\cite{p1}, we have put forward the following parametrization with 11~Kerr parameters $b_i$
\be\label{eq-m}
ds^2 &=& - \left( 1 - \frac{2 b_1 M r}{r^2 + b_2 a^2 \cos^2\theta} \right) dt^2 
\nonumber\\ &&
- \frac{4 b_3 M a r \sin^2\theta}{r^2 + b_4 a^2 \cos^2\theta} dt d\phi 
+ \frac{r^2 + b_5 a^2 \cos^2\theta}{r^2 - 2 b_6 M r + b_7 a^2} dr^2 
\nonumber\\ &&
+ \left( r^2 + b_8 a^2 \cos^2\theta \right) d\theta^2 
\nonumber\\ &&
+ \left( r^2 + b_9 a^2 + \frac{2 b_{10} M a^2 r 
\sin^2\theta}{r^2 + b_{11} a^2 \cos^2\theta} \right) \sin^2\theta d\phi^2 \, .
\ee
The line element of the Kerr metric is recovered when $b_i = 1$ for all $i$. Let us note that the metric in Eq.~(\ref{eq-m}) does not have 13~independent parameters (11~$b_i$s, $M$, and $a$), because two of them are redundant. Without the loss of generality, we can set $b_1 = 1$ (the mass measured at large distances is $b_1 M$, while $M$ cannot be determined) and $b_3 = 1$ (if $b_1 = 1$, the angular momentum measured at large distances due to frame-dragging is $b_3 M a$ while, again, $a$ cannot be measured). The number of free parameters is thus~11. However, Solar System experiments already constrain $b_6$ to be very close to 1. In the rest of the paper, we set $b_1 = b_3 = b_6 = 1$, and we allow only that the other eight Kerr parameters can be different from 1.


\section{X-ray reflection spectroscopy \label{s-r}}

Let us consider a black hole accreting from a geometrically thin and optically thick accretion disk. Within the Novikov-Thorne model~\cite{nt-model}, the disk is on the equatorial plane, orthogonal to the black hole spin. The particles of the gas in the disk follow nearly geodesic equatorial circular orbits. The inner edge of the disk is at the innermost stable circular orbit (ISCO), and the electromagnetic radiation emitted by the gas inside the disk is usually neglected. More details on the assumptions and the validity of the Novikov-Thorne model can be found, for instance, in Ref.~\cite{r2}.

A Novikov-Thorne accretion disk emits as a blackbody locally, and as a multi-color blackbody when integrated radially. The so-called corona is a hotter, usually optically-thin, electron cloud enshrouding the accretion disk. Its exact geometry is unknown. The thermal photons from the accretion disk can interact with the hot electrons in the corona. Because of inverse-Compton scattering, the corona becomes an X-ray source with a power-law spectrum. Some of these photons can illuminate the accretion disk, producing a reflection component with some fluorescence emission lines. The most prominent line is usually the iron K$\alpha$ line, which is at 6.4~keV in the case of neutral iron, and it can shift up to 6.97~keV in the case of H-like iron ions.

The iron K$\alpha$ line is very narrow in the rest-frame of the emitter, while the one observed in the spectrum of astrophysical black holes can be very broad and skewed, as a result of special and general relativistic effects (Doppler boosting, gravitational redshift, light bending) occurring in the strong gravity region of the compact object. If properly understood, a precise measurement of the shape of a broad iron line in the spectrum of an astrophysical black hole can be a powerful tool to probe its strong gravity region. As a matter of fact, one should fit the whole reflection spectrum, but the iron K$\alpha$ line is the most prominent feature and encodes most of the information on the background metric of the strong gravity region. For this reason, the technique is sometimes called the iron line method.

The study of the shape of the iron line was originally develop to measure black hole spins under the assumption that the metric around black holes is described by the Kerr solution~\cite{rey}. More recently, there have been studies to use this technique to test the nature of astrophysical black holes and verify the Kerr black hole hypothesis~\cite{ii1,ii2,ii3,ii4,ii5,i1,i2,i3,stuch,i4,i8,i9}. In the presence of high quality data and assuming to have the correct astrophysical model, the iron line method promises to provide superb constraints on possible deviations from the Kerr solution~\cite{i5,i6,i7}.

The calculations of the iron line profile in the reflection spectrum of astrophysical black holes have been already extensively discussed in the literature, see~\cite{i1,i8} and references therein. Here we employ the code described in Refs.~\cite{i1,code} with the metric in Eq.~(\ref{eq-m}). Generally speaking, the exact shape of the iron line is determined by the background metric, the inclination angle of the disk with respect to the line of sight of the observer, the geometry of the emitting region (i.e. the inner and the outer edges of the accretion disk, but here the inner edge is at the ISCO radius and the outer edge is large enough that its exact value is not important), and the intensity profile. The latter may be modeled as a simple power-law, say $I_{\rm e} \propto 1/r_{\rm e}^q$, where $I_{\rm e}$ is the local intensity, $r_{\rm e}$ is the emission radius in the disk, and $q$ is the emissivity index. A more sophisticated, but still phenomenological, model is to assume a broken power-law: $I_{\rm e} \propto 1/r_{\rm e}^{q_1}$ for $r_{\rm e} < r_{\rm b}$ and $I_{\rm e} \propto 1/r_{\rm e}^{q_2}$ for $r_{\rm e} > r_{\rm b}$, and we have three parameters, namely $q_1$, $q_2$, and the breaking radius $r_{\rm b}$.

\begin{figure*}
\vspace{0.4cm}
\begin{center}
\includegraphics[type=pdf,ext=.pdf,read=.pdf,width=8.0cm]{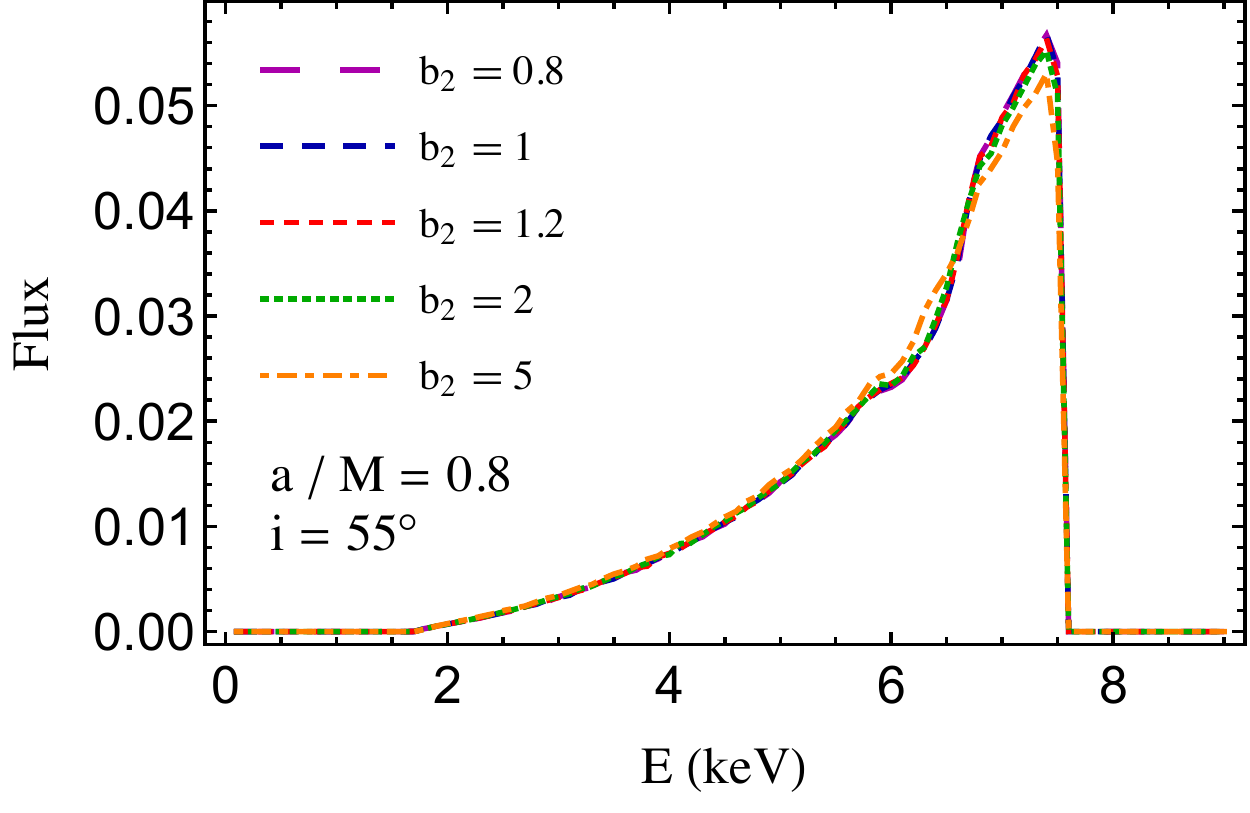}
\hspace{0.8cm}
\includegraphics[type=pdf,ext=.pdf,read=.pdf,width=8.0cm]{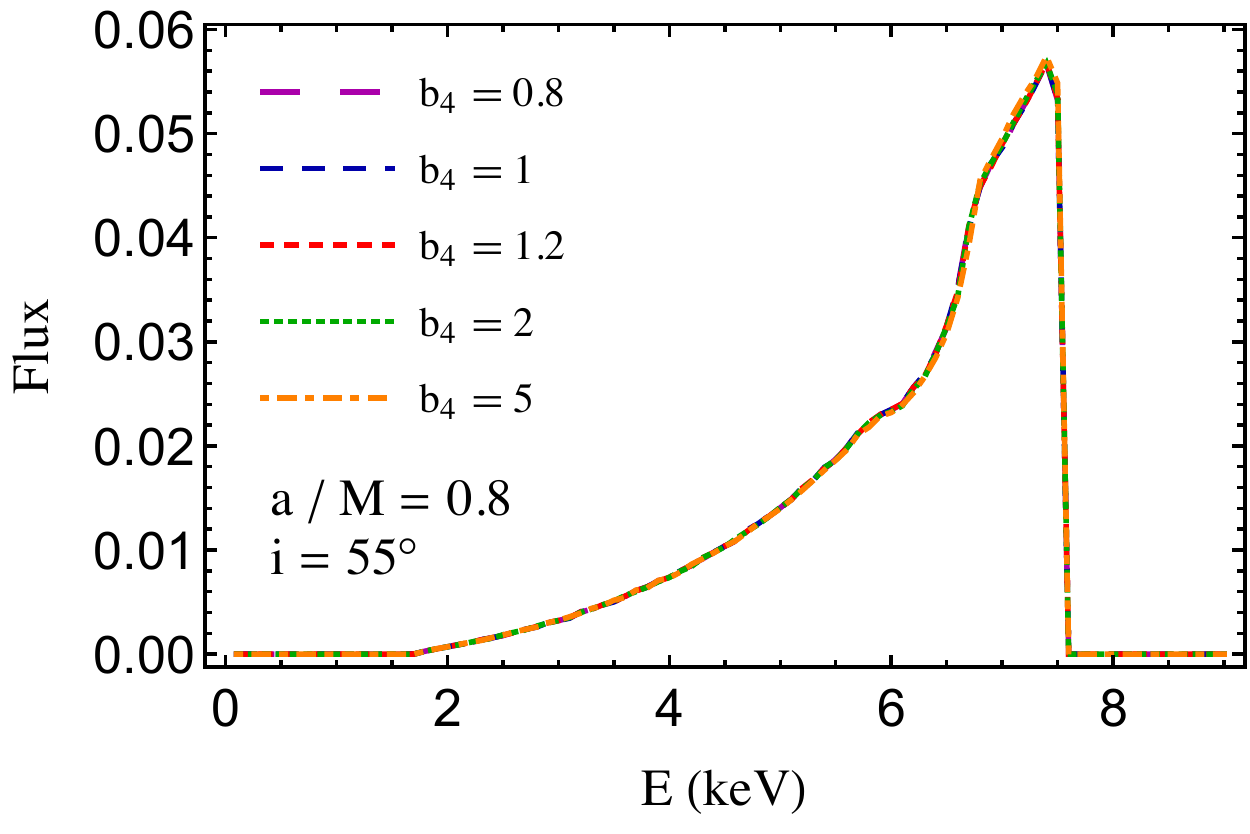}\\
\vspace{0.8cm}
\includegraphics[type=pdf,ext=.pdf,read=.pdf,width=8.0cm]{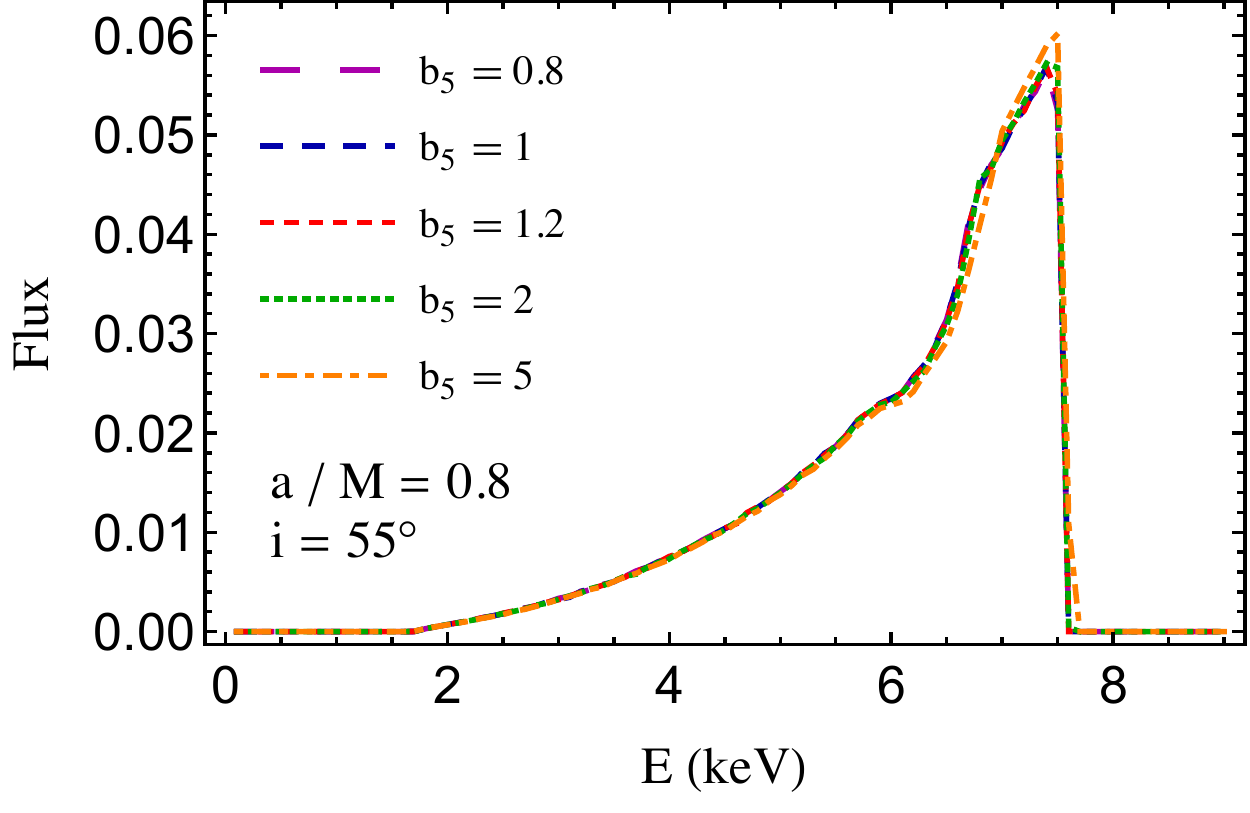}
\hspace{0.8cm}
\includegraphics[type=pdf,ext=.pdf,read=.pdf,width=8.0cm]{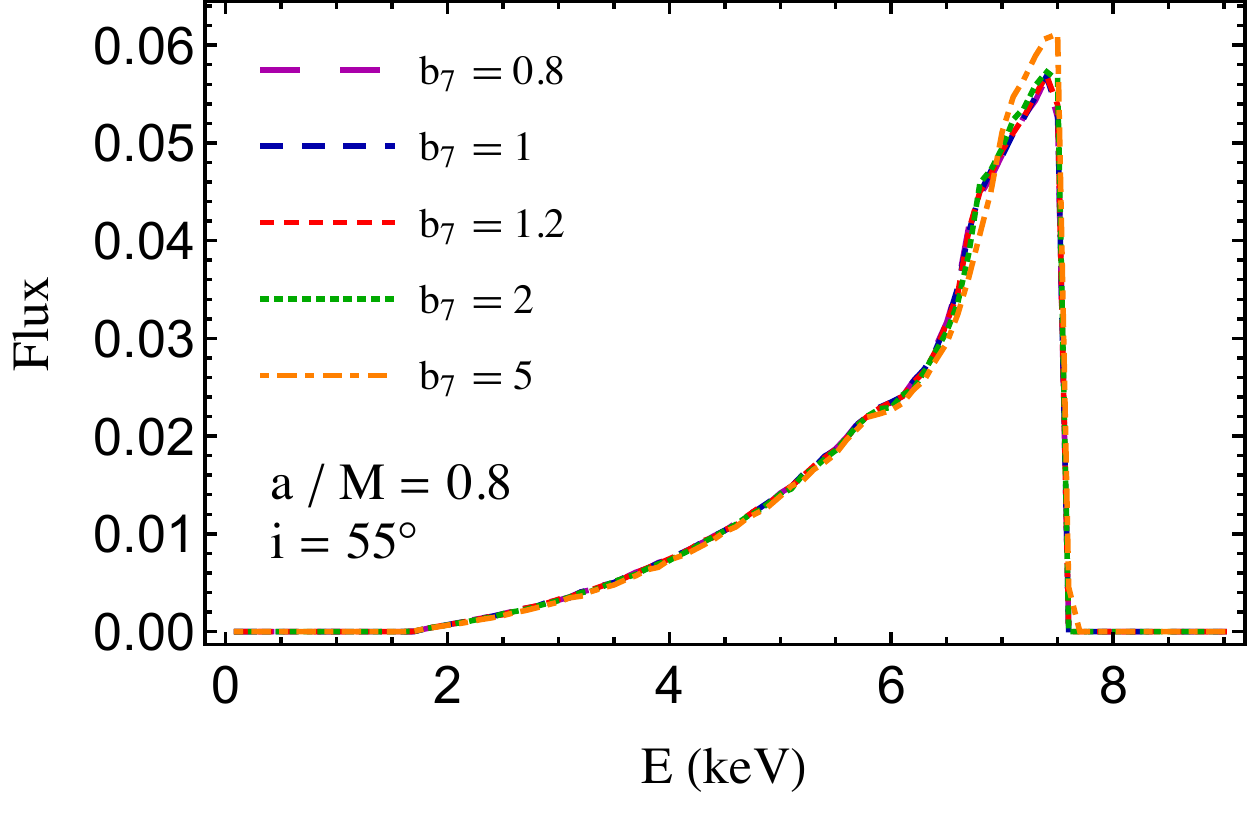}\\
\vspace{0.8cm}
\includegraphics[type=pdf,ext=.pdf,read=.pdf,width=8.0cm]{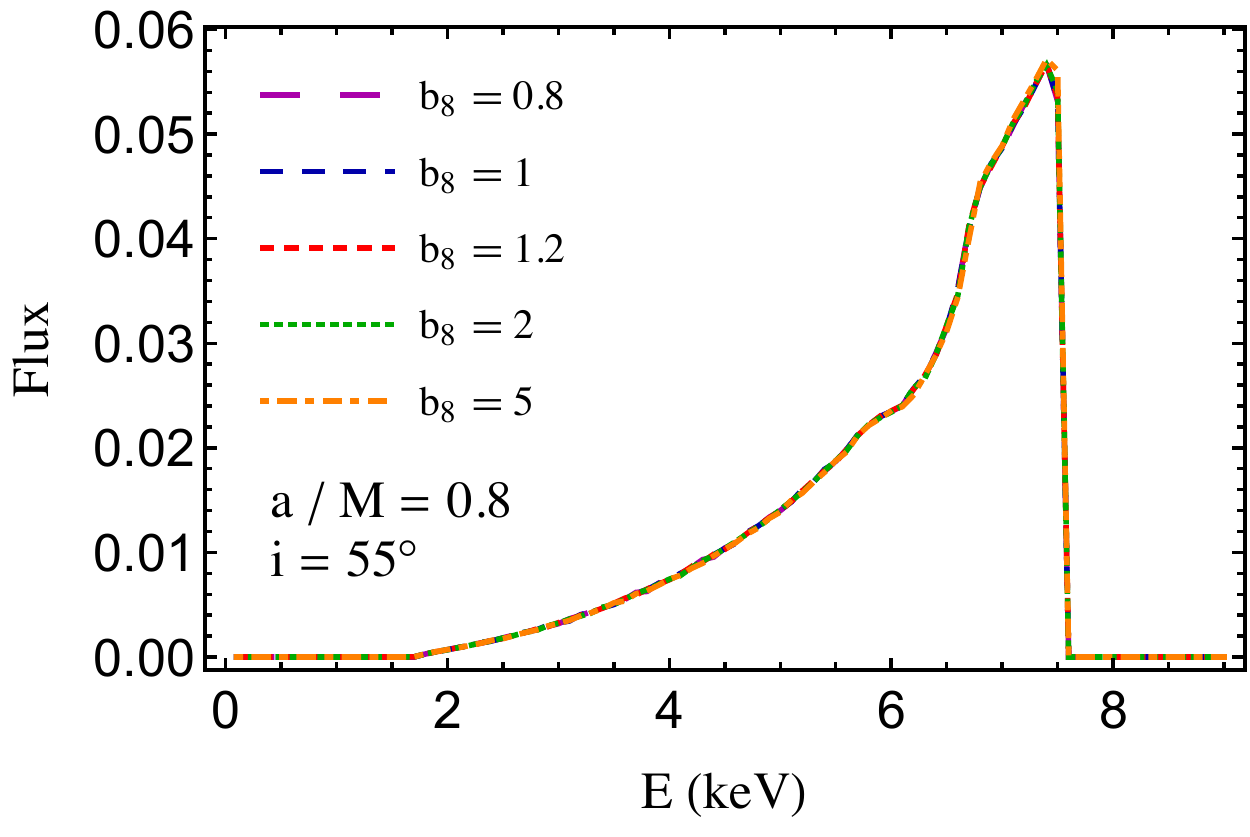}
\hspace{0.8cm}
\includegraphics[type=pdf,ext=.pdf,read=.pdf,width=8.0cm]{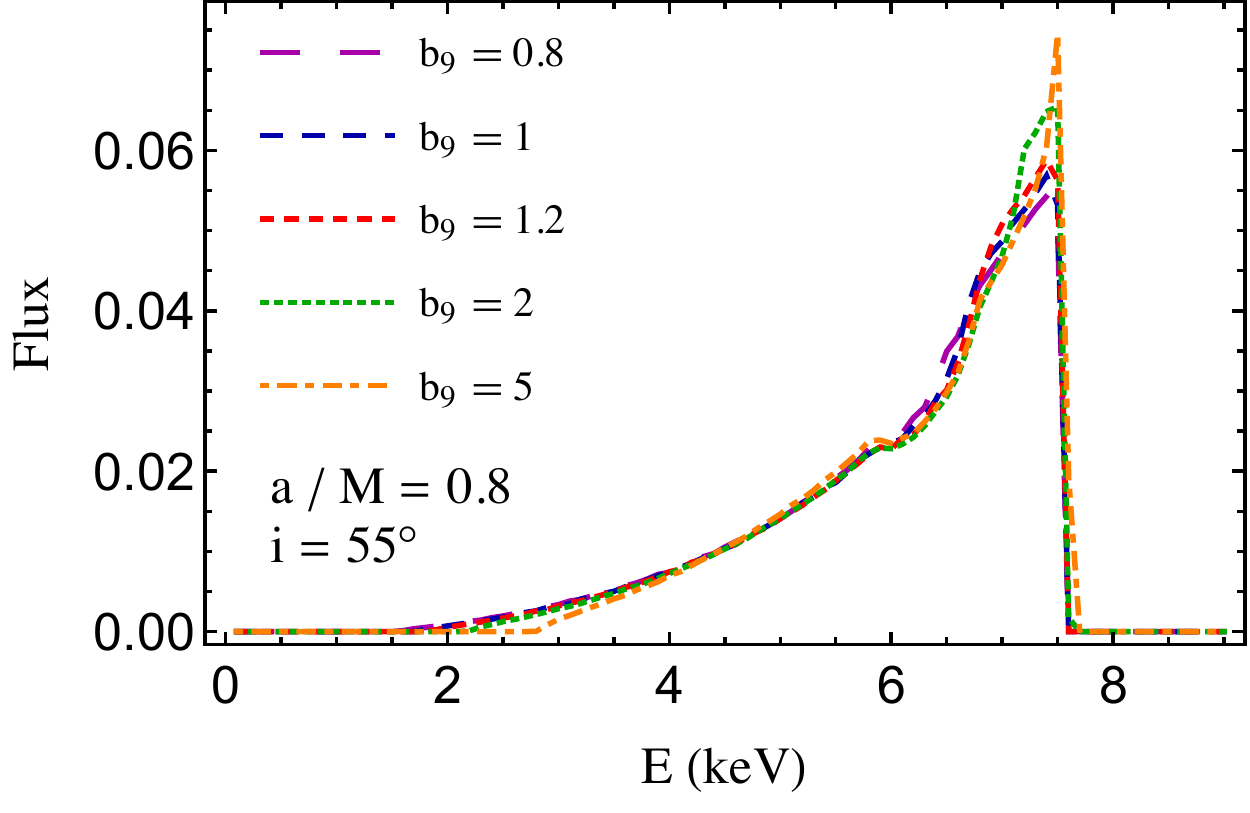} 
\end{center}
\vspace{-0.3cm}
\caption{Impact of the parameters $b_2$ (top left panel), $b_4$ (top right panel), $b_5$ (central left panel), $b_7$ (central right panel), $b_8$ (bottom left panel), and $b_9$ (bottom right panel) on the shape of the iron line. The flux of the photon number is in arbitrary units. See the text for more details. \label{fig1}}
\end{figure*}

\begin{figure*}
\vspace{0.4cm}
\begin{center}
\includegraphics[type=pdf,ext=.pdf,read=.pdf,width=8.0cm]{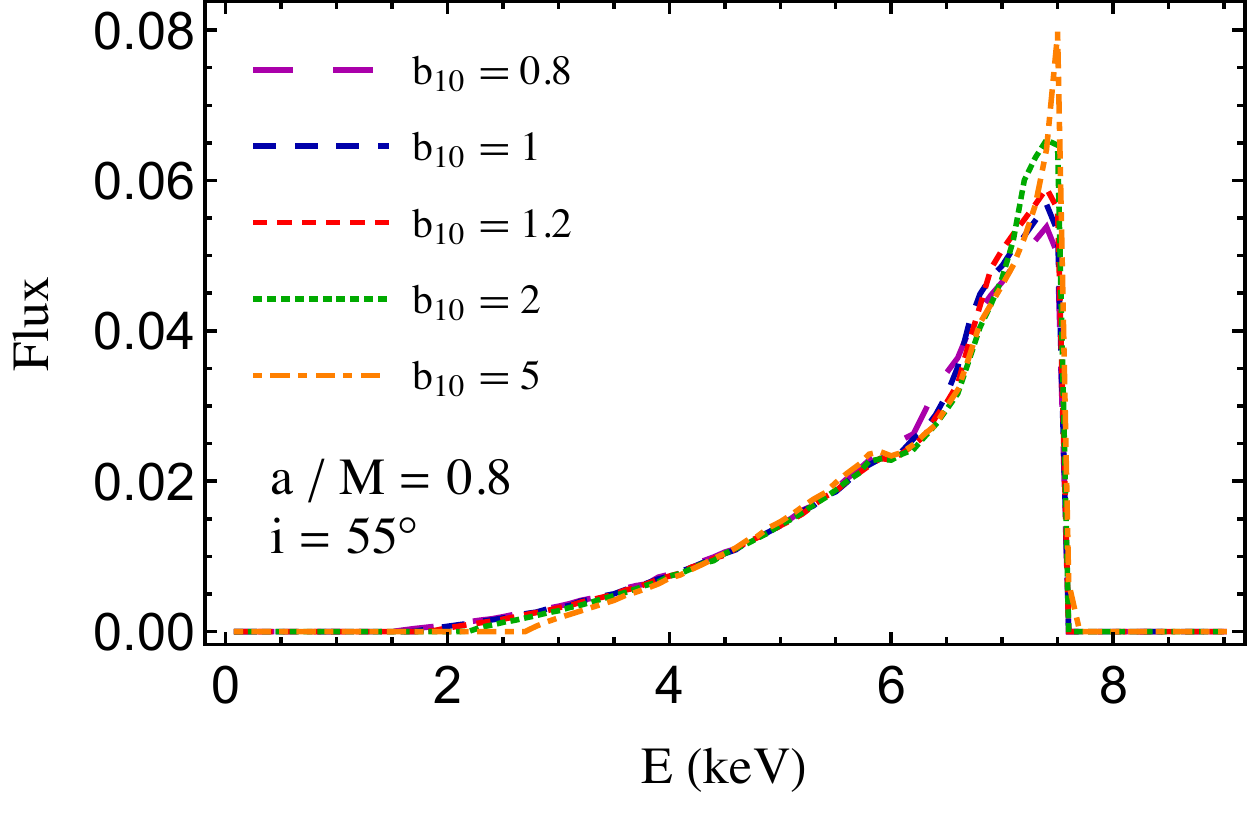}
\hspace{0.8cm}
\includegraphics[type=pdf,ext=.pdf,read=.pdf,width=8.0cm]{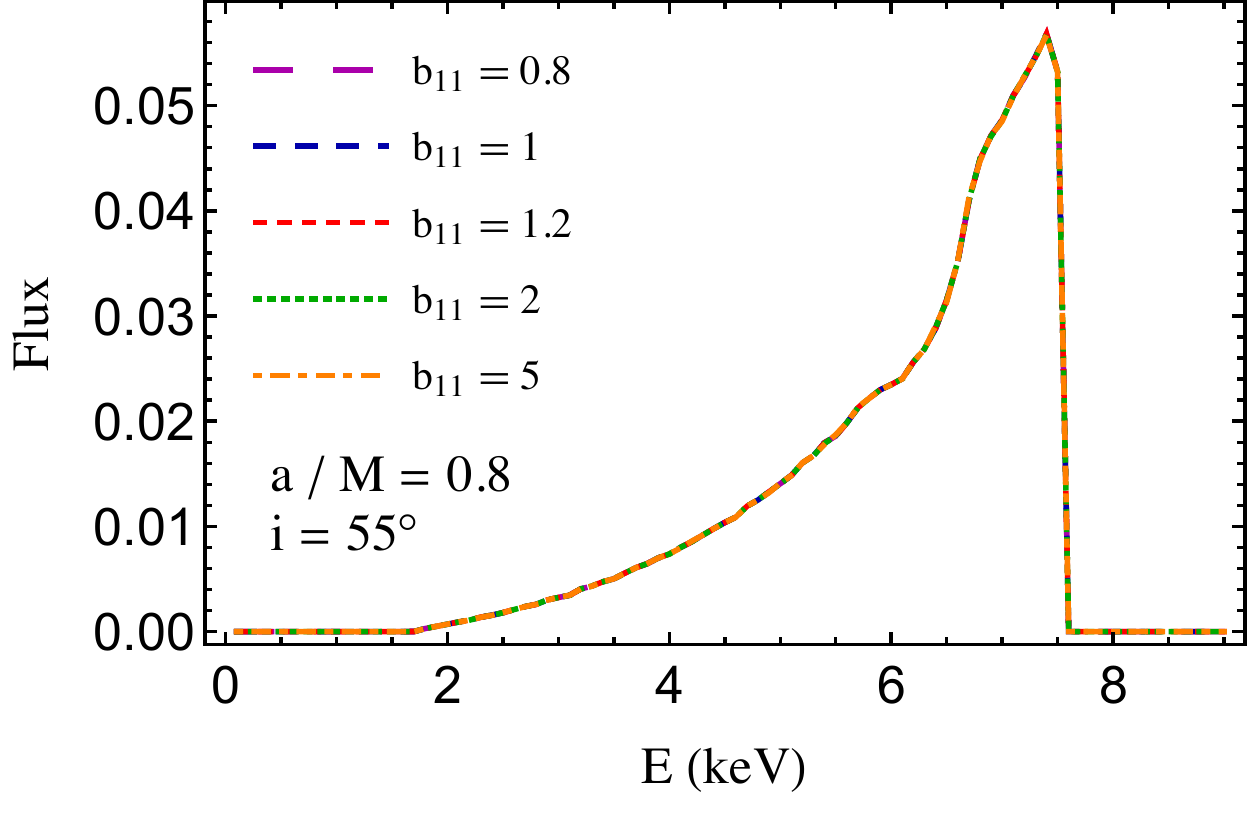}
\end{center}
\vspace{-0.3cm}
\caption{As in Fig.~\ref{fig1} for the parameters $b_{10}$ (left panel) and $b_{11}$ (right panel). \label{fig2}}
\vspace{-0.7cm}
\begin{center}
\includegraphics[type=pdf,ext=.pdf,read=.pdf,width=15.0cm]{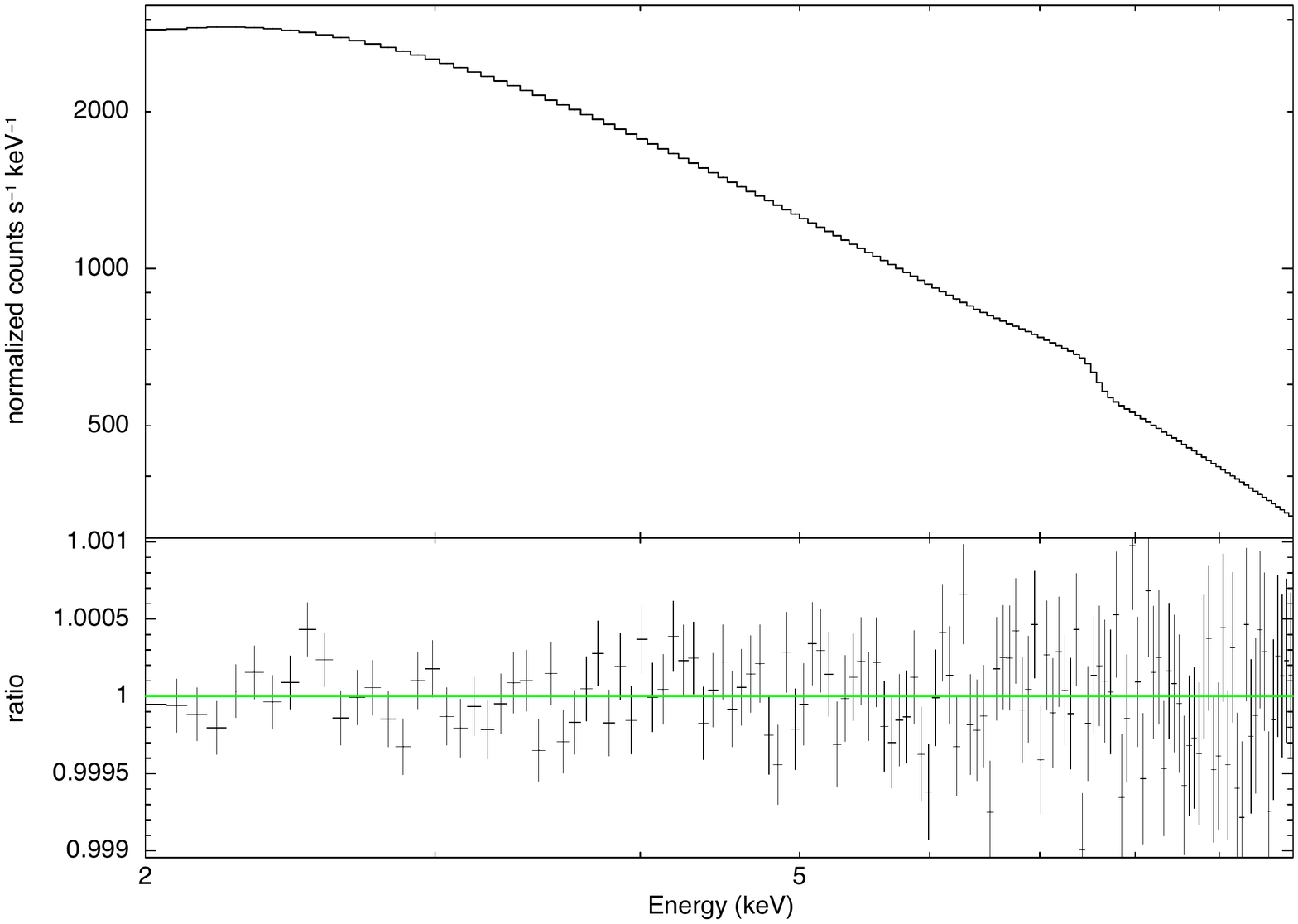}
\end{center}
\vspace{-1.2cm}
\caption{Folded spectrum (top panel) and ratio between the simulated data and the best fit (bottom panel) For the simulation of a Kerr black hole with spin parameter $a_* = 0.8$ and observed from a viewing angle $i = 55^\circ$. The minimum of the reduced $\chi^2$ is about 0.99. See the text for more details. \label{fig3-k}}
\end{figure*}

Figs.~\ref{fig1} and \ref{fig2} show the iron lines expected in the family of spacetimes described by Eq.~(\ref{eq-m}). In all plots, the spin parameter is $a_* = 0.8$ and the inclination angle of the disk with respect to the line of sight of the observer is $i = 55^\circ$. The emissivity profile is assumed to be $I_{\rm e} \propto 1/r_{\rm e}^3$, which corresponds to the Newtonian limit at larger radii in the lamppost corona geometry. From these plots, without a qualitative analysis, we see that the impact of $b_{11}$ on the iron line shape is very weak, and it is thus presumably difficult to measure its value. It seems that all the lines overlap in the plot. For the Kerr parameters $b_4$ and $b_8$, the impact on the iron line is still weak, but there are some differences if we change their value. $b_2$, $b_5$, $b_7$, $b_9$, and $b_{10}$ seem to be able to produce stronger effects. In the next section, we will perform a simple quantitative analysis with some simulations.


\begin{figure*}
\vspace{-0.3cm}
\begin{center}
\includegraphics[type=pdf,ext=.pdf,read=.pdf,width=9.2cm]{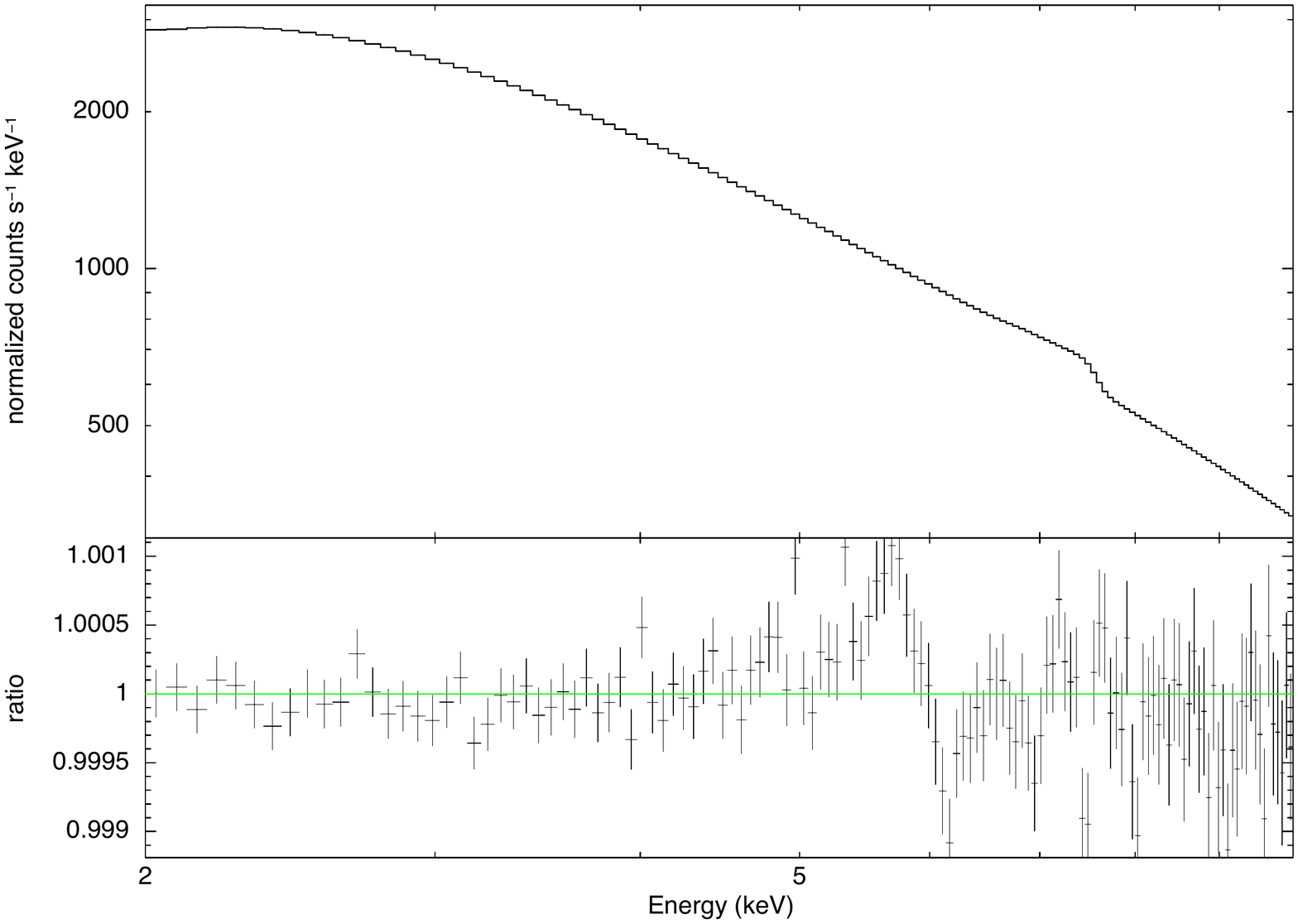}
\hspace{-0.8cm}
\includegraphics[type=pdf,ext=.pdf,read=.pdf,width=9.2cm]{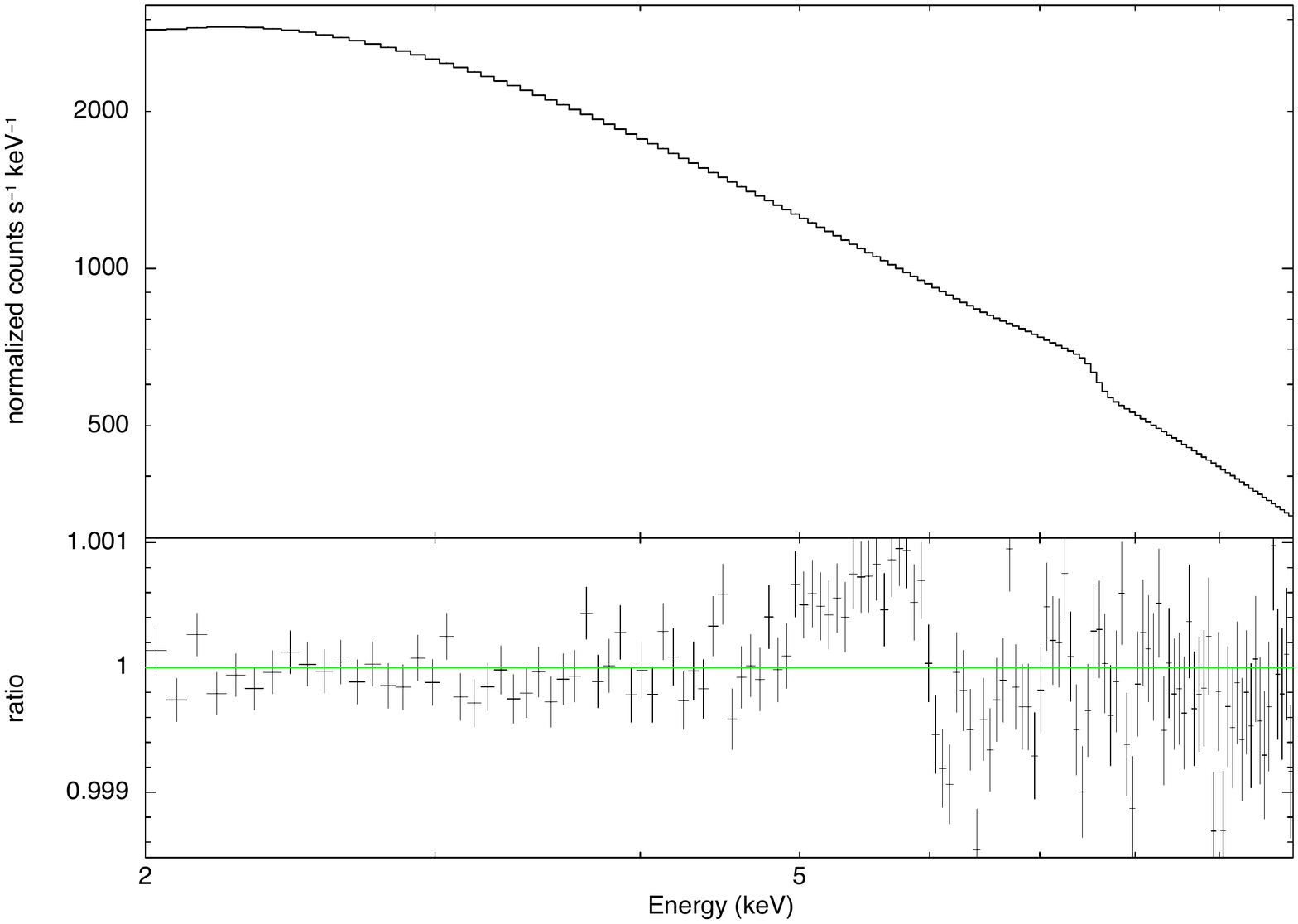} \\
\vspace{-0.4cm}
\includegraphics[type=pdf,ext=.pdf,read=.pdf,width=9.2cm]{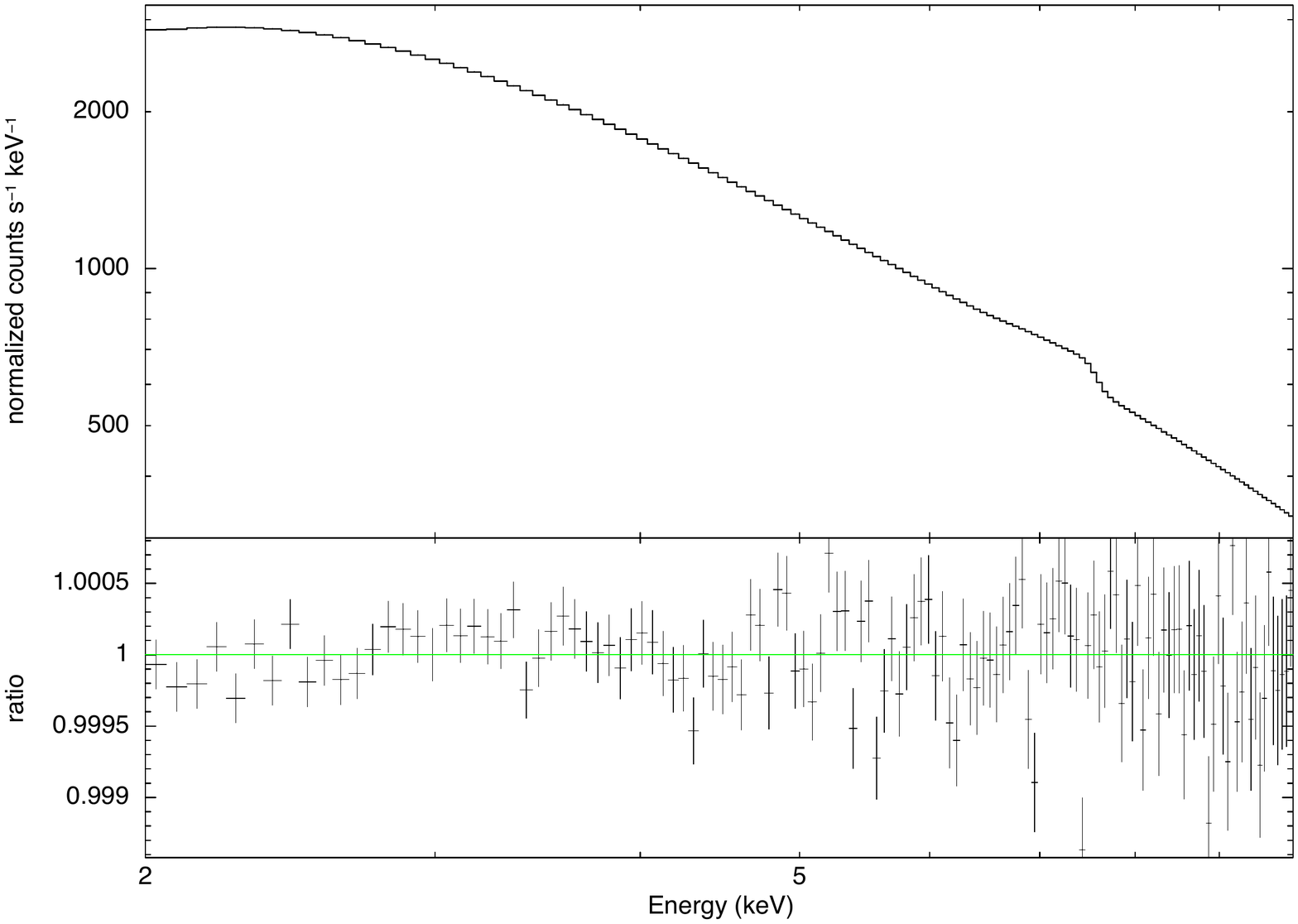}
\hspace{-0.8cm}
\includegraphics[type=pdf,ext=.pdf,read=.pdf,width=9.2cm]{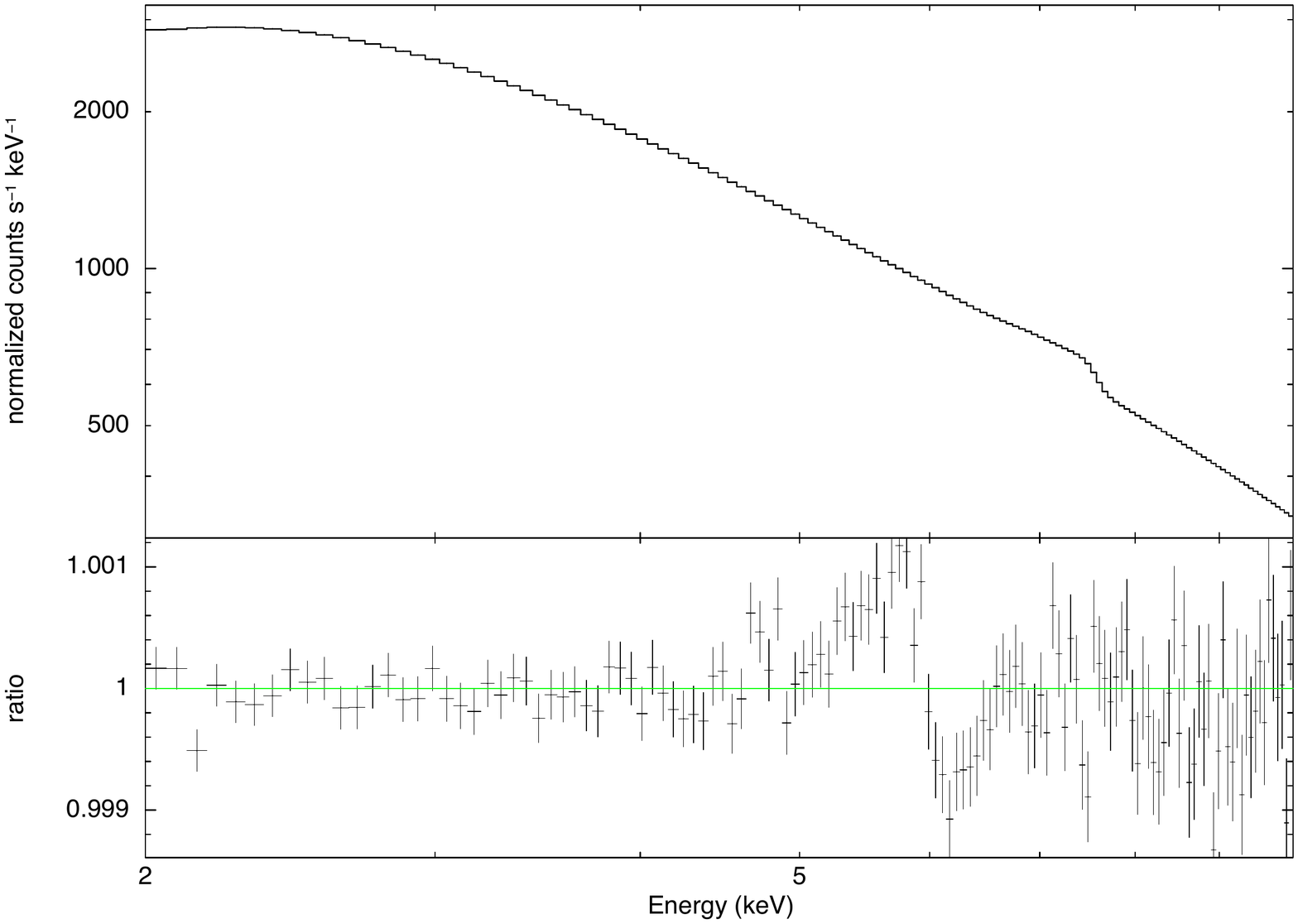}
\end{center}
\vspace{-0.4cm}
\caption{Top left panel: As in Fig.~\ref{fig3-k} for a background metric with $b_2 = 1.2$ and all other Kerr parameters set to 1. The minimum of the reduced $\chi^2$ is about 1.63. Top right panel: As in Fig.~\ref{fig3-k} for a background metric with $b_4 = 1.2$ and all other Kerr parameters set to 1. The minimum of the reduced $\chi^2$ is about 2.16. Bottom left panel: As in Fig.~\ref{fig3-k} for a background metric with $b_5 = 1.2$ and all other Kerr parameters set to 1. The minimum of the reduced $\chi^2$ is about 1.16. Bottom right panel: As in Fig.~\ref{fig3-k} for a background metric with $b_7 = 1.2$ and all other Kerr parameters set to 1. The minimum of the reduced $\chi^2$ is about 1.90. See the text for more details. \label{fig3-b2457}}
\end{figure*}

\begin{figure*}
\vspace{-0.3cm}
\begin{center}
\includegraphics[type=pdf,ext=.pdf,read=.pdf,width=9.2cm]{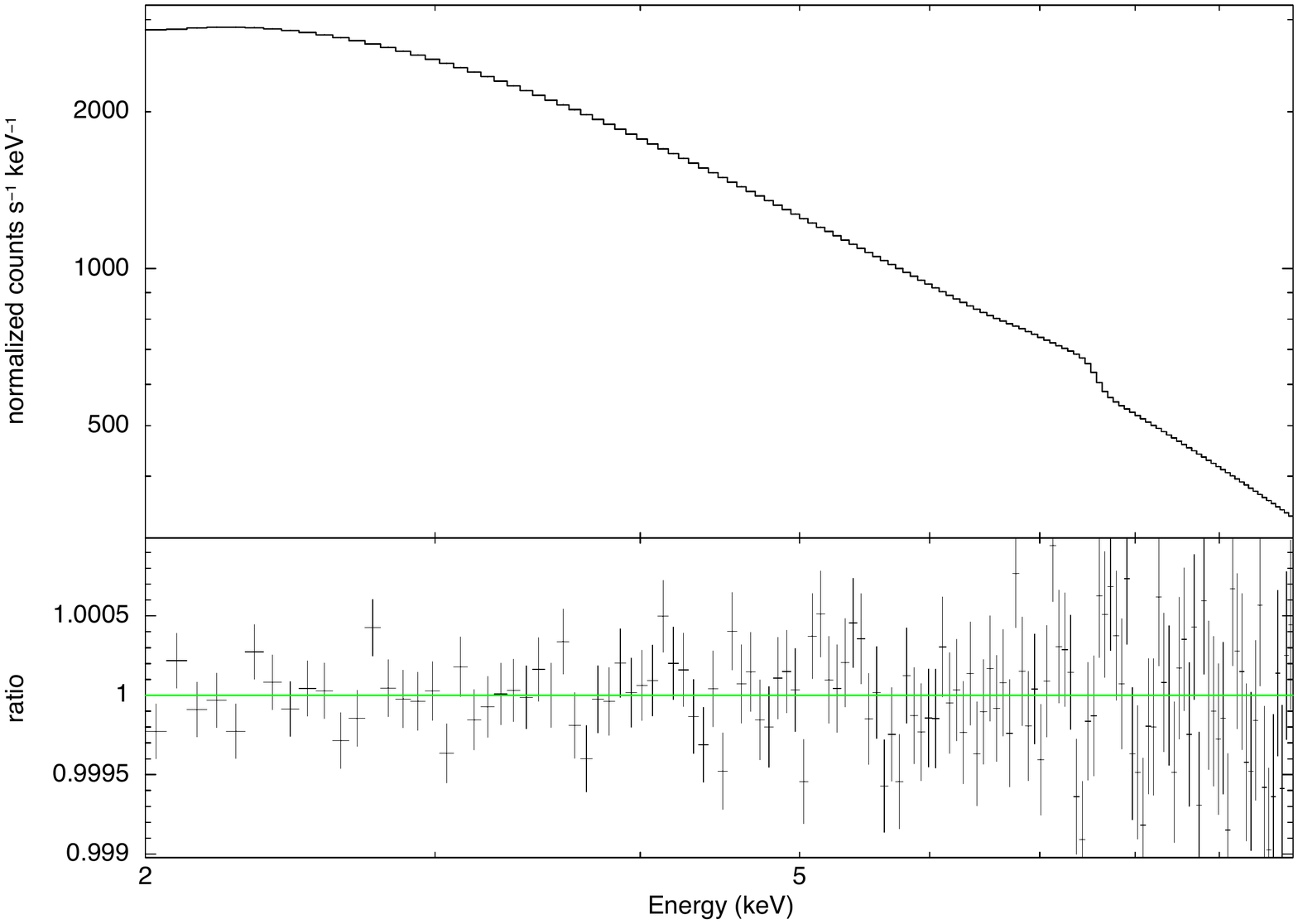}
\hspace{-0.8cm}
\includegraphics[type=pdf,ext=.pdf,read=.pdf,width=9.2cm]{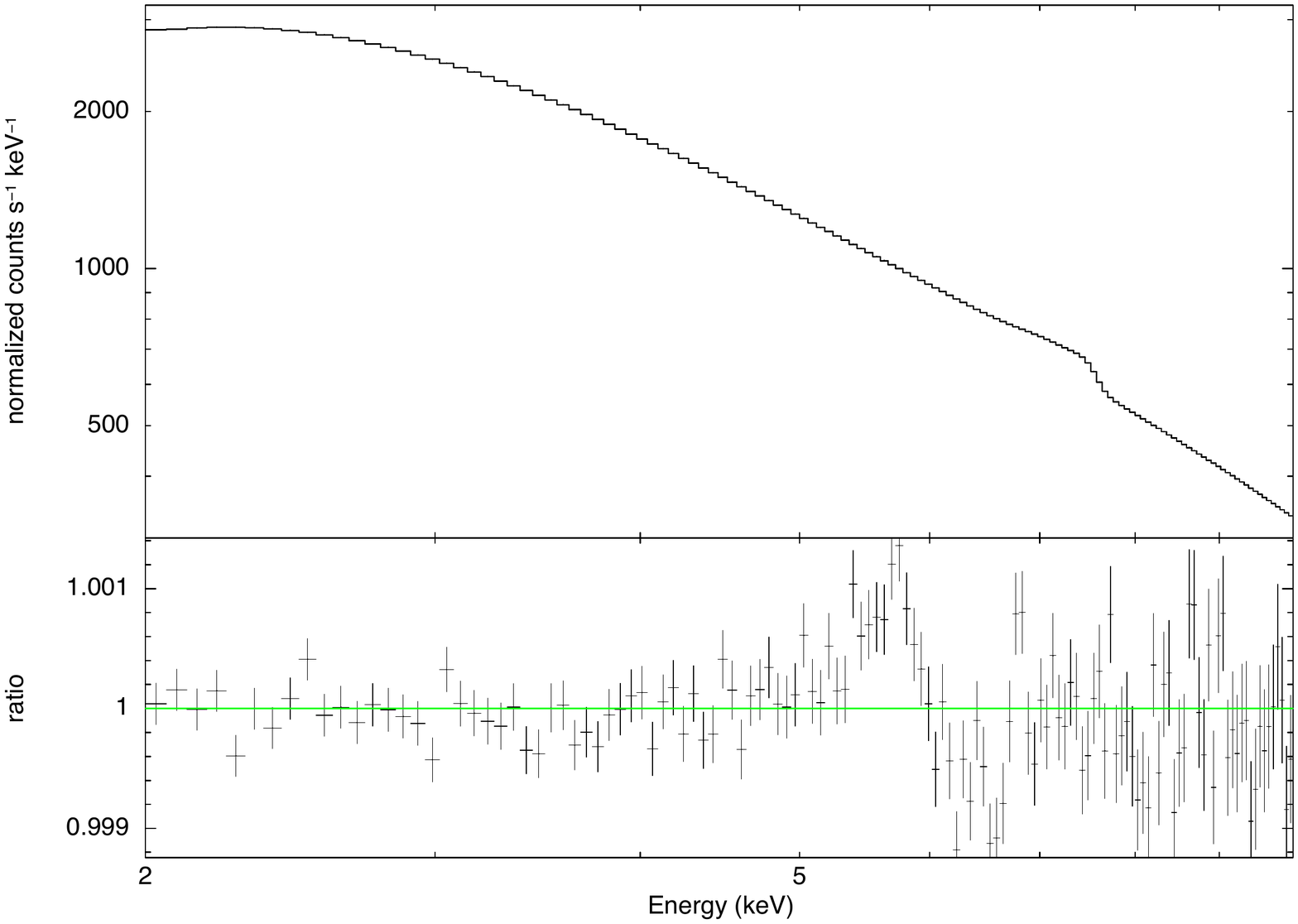} \\
\vspace{-0.4cm}
\includegraphics[type=pdf,ext=.pdf,read=.pdf,width=9.2cm]{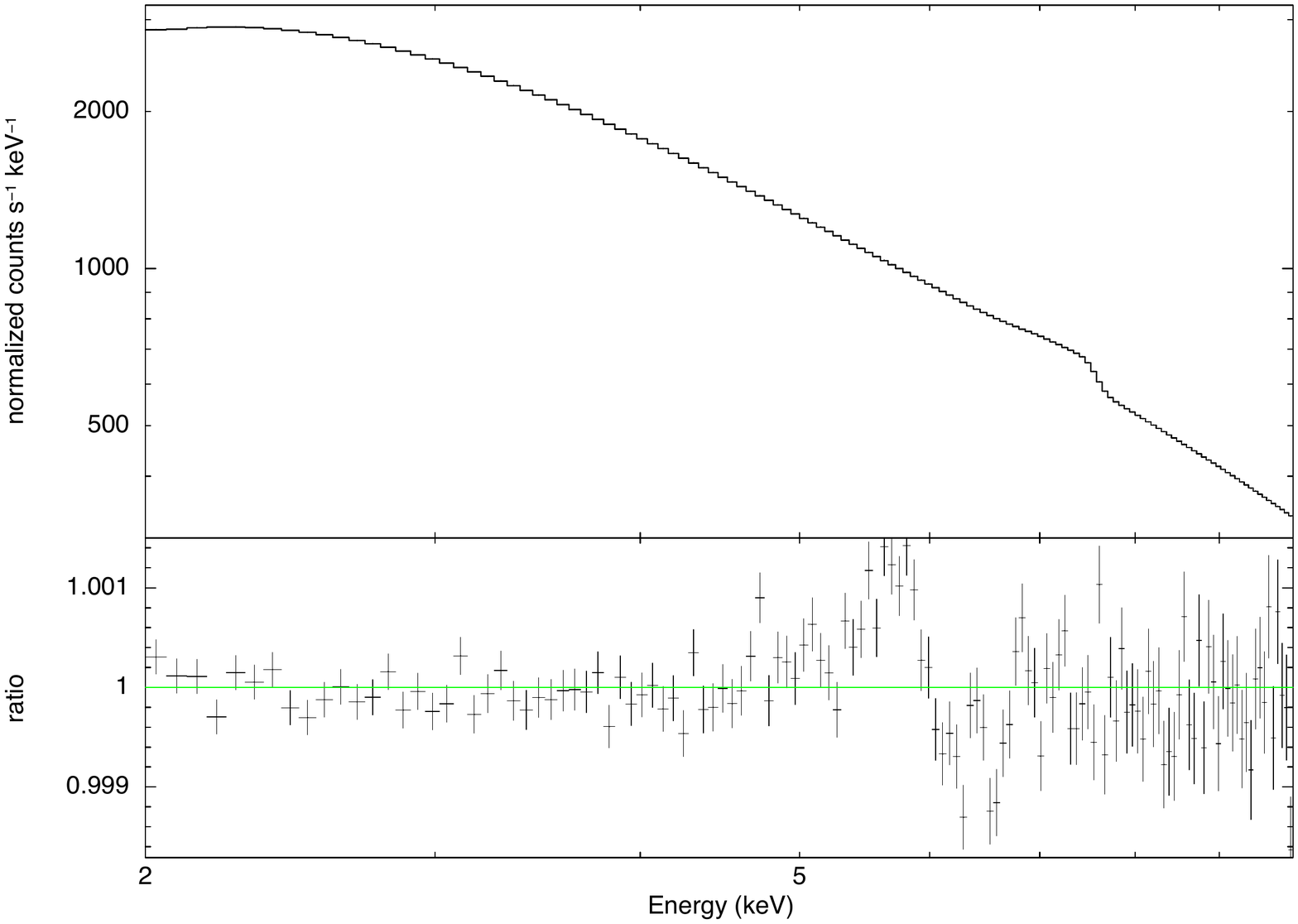}
\hspace{-0.8cm}
\includegraphics[type=pdf,ext=.pdf,read=.pdf,width=9.2cm]{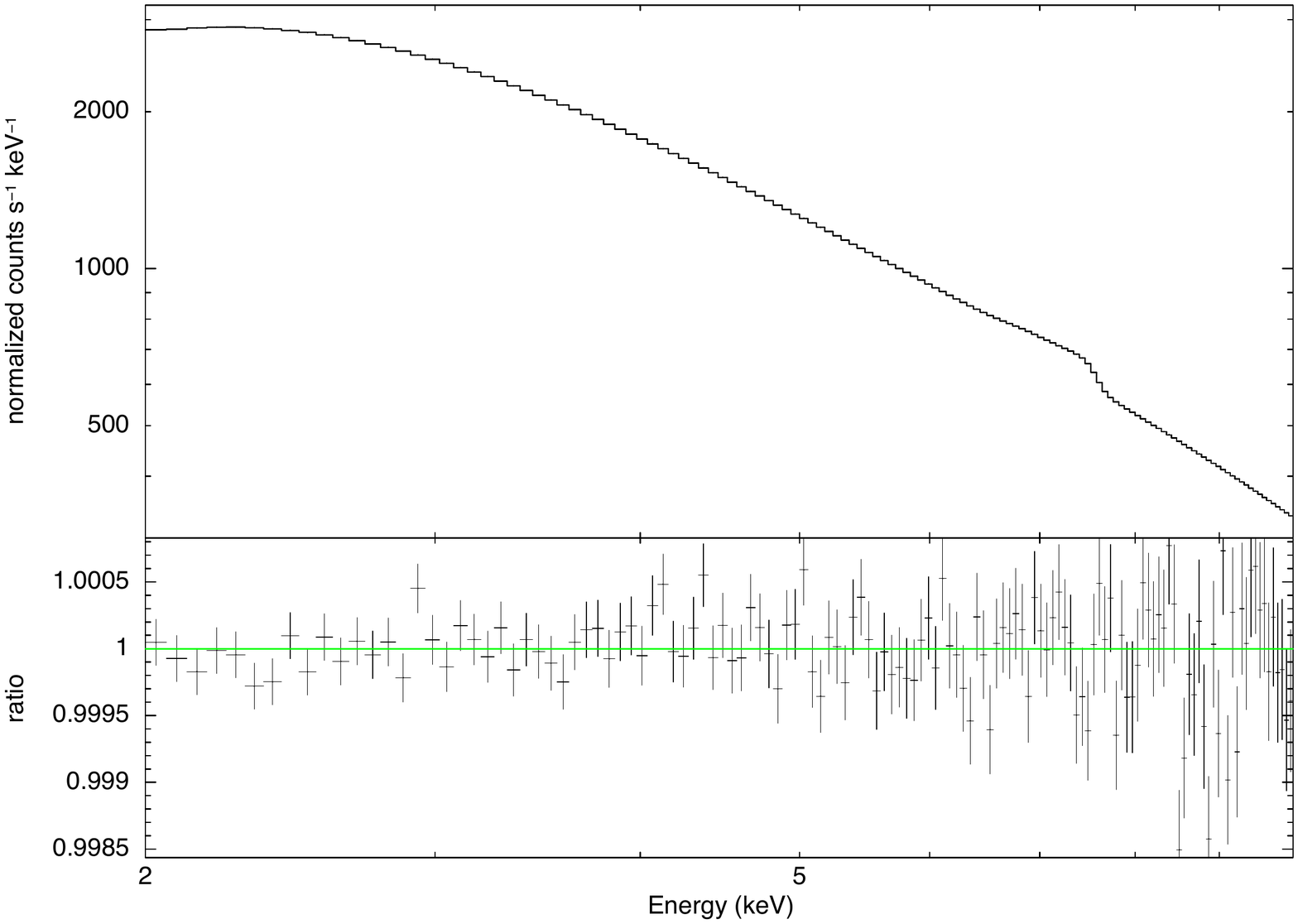}
\end{center}
\vspace{-0.4cm}
\caption{Top left panel: As in Fig.~\ref{fig3-k} for a background metric with $b_8 = 1.2$ and all other Kerr parameters set to 1. The minimum of the reduced $\chi^2$ is about 1.17. Top right panel: As in Fig.~\ref{fig3-k} for a background metric with $b_9 = 1.2$ and all other Kerr parameters set to 1. The minimum of the reduced $\chi^2$ is about 2.19. Bottom left panel: As in Fig.~\ref{fig3-k} for a background metric with $b_{10} = 1.2$ and all other Kerr parameters set to 1. The minimum of the reduced $\chi^2$ is about 2.57. Bottom right panel: As in Fig.~\ref{fig3-k} for a background metric with $b_{11} = 5$ and all other Kerr parameters set to 1. The minimum of the reduced $\chi^2$ is about 1.06. See the text for more details. \label{fig3-b891011}}
\end{figure*}

\section{Simulations \label{s-s}}

In this section we want to understand if the analysis of the reflection spectrum of a black hole can potentially constrain the Kerr parameters $b_2$, $b_4$, $b_5$, $b_7$, $b_8$, $b_9$, $b_{10}$, and $b_{11}$. Our analysis is similar to those presented in Refs.~\cite{f1,f2,f3}.

We simulate observations of a bright black hole binary. The reference spectrum is a power-law with photon index $\Gamma = 2$, which describes the direct component from the corona, plus a single iron K$\alpha$ line, describing the reflection spectrum. The iron line is computed with our code and we allow that one of the Kerr parameters is different from 1. In all the simulations, the equivalent width of the iron line is about 200~eV, and the photon flux in the range 2-10~keV is about $2.6 \cdot 10^{-9}$~erg/cm$^2$/s, which corresponds to the photon flux of a bright black hole binary. The observations are simulated with the LAD instrument, on board of the X-ray mission eXTP, which is currently scheduled to be launched in 2022~\cite{extp}. The exposure time is 200~ks.

We use Xspec\footnote{https://heasarc.gsfc.nasa.gov/xanadu/xspec/} with the redistribution matrix, ancillary, and background files of LAD/eXTP following the forward-folding approach common in X-ray astronomy. The actual spectrum measured by an instrument is given as a photon count per spectral bin and can be written as
\be\label{eq-ffa}
C(h) = \tau \int R(h,E) \, A(E) \, s(E) \, dE \, ,
\ee
where $h$ is the spectral channel, $\tau$ is the exposure time, $E$ is the photon energy, $R(h,E)$ is the redistribution matrix (essentially the response of the instrument), $A(E)$ is the effective area (say the efficiency of the instrument, which is given in the ancillary file), and $s(E)$ is the intrinsic spectrum of the source. In general, the redistribution matrix cannot be inverted, and for this reason we have to deal with $C(h)$ (folded spectrum). With the forward-folding approach, we consider a set of parameter values for the intrinsic spectrum, we find $C(h)$, we compare the ``observed spectrum'' with the folded spectrum with some goodness-of-fit statistical test (e.g. $\chi^2$), and we repeat all these steps to find ``the best fit'' by changing the input parameters in the theoretical model. In our case, the observed spectrum is not from real data but from a simulation. It is obtained assuming a certain intrinsic spectrum $s(E)$ and including the background in Eq.~(\ref{eq-ffa}). The latter includes the noises of the instrument and of the environment (e.g. photons not from the target source or cosmic rays). The observational error has also the intrinsic noise of the source (Poisson noise) due to the fact the spectrum is as a photon count per bin and is not a continuous quantity. Thanks to the unprecedented large effective area of LAD/eXTP, which will be about 35,000~cm$^2$ at 6~keV, the Poisson noise is very low in our case.

These simulated observations are fitted with a Kerr model. We use Xspec and fit the simulated data with a power-law and a Kerr iron line. The latter is modeled with RELLINE~\cite{relline}. We have eight free parameters: the photon index of the power-law $\Gamma$, the normalization of the power-law component, the spin of the black hole $a_*$, the viewing angle $i$, the two indices $q_1$ and $q_2$ of the broken power-law of the emissivity profile, the breaking radius $r_{\rm b}$, and the normalization of the iron line.

The fit of a simulated observation of a Kerr black hole (i.e. $b_i = 1$ for all $i$) is shown in Fig.~\ref{fig3-k}. The Kerr black hole has the spin parameter $a_* = 0.8$, the viewing angle is $i = 55^\circ$, and the intensity profile of the reflection spectrum is modeled with a simple power-law whose emissivity index is $q = 3$. The fit is good, as it should be because both the simulation and the fit assume the Kerr metric, but it is a necessary check because the simulations are done with our code and the fit is done with RELLINE.

Figs.~\ref{fig3-b2457} and \ref{fig3-b891011} show the fits of the simulated observations in which one of the $b_i$ is different from 1. As in the previous simulation, $a_* = 0.8$, $i = 55^\circ$, and $I_{\rm e} \propto 1/r^3_{\rm e}$. In every figure, the top panel is the folded spectrum and the bottom panel is the ratio between the simulated data and the best fit. From the latter, it is possible to say whether a fit is good or bad. If there are clear unresolved features, the model used to fit the data is not adequate. In our case, it means that an iron line of a Kerr spacetime cannot provide a good fit, and therefore we can constrain the Kerr parameter. The reduced $\chi^2$ of the best fit may also be used to evaluate the quality of a fit, but, in general, it depends on the specific simulation.

Figs.~\ref{fig3-b2457} and \ref{fig3-b891011} are for $b_2$, $b_4$, $b_5$, $b_7$, $b_8$, $b_9$, $b_{10}$, and $b_{11}$. In every figure, the Kerr parameter under investigation has the value 1.2 (5 for $b_{11}$) and all the other parameters are set to 1. There are clear unresolved features in the ratio panels for $b_2$, $b_4$, $b_7$, $b_9$, and $b_{10}$. This means that a 200~ks observation of a bright black hole binary with LAD/eXTP can confirm whether these parameters are 1 at the level of 20\% or better. In the case of $b_5$ and $b_8$, the fit is not too bad. There are not clear unresolved features in the ratio between the simulated data and the best fit. The minimum of the reduced $\chi^2$ is also acceptable. We have checked that the fit gets worse if either we increase the value of these Kerr parameter or we consider a longer exposure time. In other words, even $b_5$ and $b_8$ can be constrained by LAD/eXTP, but the constraints are weaker than those on the other Kerr parameters. From the iron lines in Figs.~\ref{fig1} and \ref{fig2}, without a quantitative analysis, we had seen that variations of the values of $b_2$, $b_5$, $b_7$, $b_9$, and $b_{10}$ could produce larger differences in the iron line than variations of the values of $b_4$ and $b_8$, but in those plots all other parameters were fixed. In Figs.~\ref{fig3-b2457} and \ref{fig3-b891011}, we fit the simulations with eight free parameters ($\Gamma$, the normalization of the power-law component, $a_*$, $i$, $q_1$, $q_2$, $r_{\rm b}$, and the normalization of the iron line) and therefore we can figure out possible correlations in their estimate. We find that $b_5$ and $b_8$ are more difficult to constrain than $b_2$, $b_4$, $b_7$, $b_9$, and $b_{10}$.

A special case is the parameter $b_{11}$. From the right panel in Fig.~\ref{fig2}, we do not see any appreciable variation in the iron line by changing the value of $b_{11}$. The fit of the simulated observation in the bottom left panel in Fig.~\ref{fig3-b891011} confirms that $b_{11}$ has a very weak impact on the iron line and is very difficult to constrain from observations. In this simulation $b_{11} = 5$. Despite such a large deviation from the Kerr value, the fit is good. We thus conclude that $b_{11}$ may not be constrained with the analysis of the reflection spectrum.


\section{Summary and conclusions \label{s-c}}

In Ref.~\cite{p1}, we have proposed a new parametrization to test the metric around astrophysical black holes. It is widely believed that the spacetime geometry around astrophysical black holes is well approximated by the Kerr metric of general relativity, but macroscopic deviations may be possible in the presence of new physics. In our proposal, we have a number of Kerr parameters. They are all 1 in the Kerr case, and may be different from 1 in the presence of new physics.

In Ref.~\cite{p1}, we had considered the possibility of constraining the Kerr parameters from an accurate observation of the apparent photon capture sphere, a kind of measurement that may be possible in the near future but on which there are currently many uncertainties. We found that the parameters $b_4$, $b_5$, $b_7$, and $b_{11}$ have a very weak impact on the shape of the apparent photon capture sphere, and this suggests that it is very challenging to constrain their value from a putative accurate detection of the apparent photon capture sphere of a black hole.

In the present paper, we have studied the possibility of constraining the values of these Kerr parameters via the X-ray reflection spectroscopy. We have simulated 200~ks observations with LAD/eXTP, assuming that the Kerr parameters may have a value different from 1. We have fitted these simulations with Kerr models to see whether it is possible to find a good fit. If the fit is good, the impact of the Kerr parameter under study is too weak and it cannot produce appreciable deviations from an iron line in the Kerr spacetime. If the fit is bad, the data cannot be fitted with an iron line in a Kerr spacetime and therefore it is possible to constrain the Kerr parameter.

Our results look quite promising, in the sense that all the Kerr parameters -- with the remarkable exception of $b_{11}$ -- produce noticeable effects in the shape of the iron line profile and can thus be tested by a 200~ks observation with LAD/eXTP. In the case of $b_2$, $b_4$, $b_7$, $b_9$, and $b_{10}$, Kerr models cannot provide a good fit already if one of these parameters is 1.2. For $b_5$ and $b_8$ equal to 1.2, a 200~ks observation with LAD/eXTP is marginally consistent with a Kerr model, but we find that the fit is not acceptable as the value of the Kerr parameter increases.


\begin{acknowledgments}
M.G.-N. thanks the School of Astronomy at the Institute for Research in Fundamental Sciences (IPM), Tehran, Iran, where most of this work was done. M.G.-N. and C.B. were supported by the NSFC (grants 11305038 and U1531117) and the Thousand Young Talents Program. M.G.-N. also acknowledges support from the China Scholarship Council (CSC), grant No.~2014GXZY08. C.B. also acknowledges support from the Alexander von Humboldt Foundation.
\end{acknowledgments}


\end{document}